\documentclass[]{aastex}

\usepackage{emulateapj5}
\usepackage{onecolfloat}
\usepackage{graphicx} 
\usepackage{fancyheadings} 
\usepackage{ulem}
\usepackage{rotating}
\usepackage{lscape}

\newcommand{\grbnm}{HG~031203}

\newcommand{\msol}{M_\odot}

\newcommand{\halph}{H$\alpha$}

\newcommand{\hbeta}{H$\beta$}
\newcommand{\hgamma}{H$\gamma$}
\newcommand{\hdelta}{H$\delta$}

\newcommand{\cm}[1]{\, {\rm cm^{#1}}}

\newcommand{\sci}[1]{{\rm \; \times \; 10^{#1}}}

\newcommand{\mkms}{{\rm \; km\;s^{-1}}}
\newcommand{\tskip}{\tablevspace{1pt}}

\begin{document}

\twocolumn[%
\submitted{Accepted to the Astrophysical Journal: April 20, 2004}
\title{The Host Galaxy of GRB~031203: 
Implications of its low metallicity, low redshift, and starburst nature}

\author{ Jason X. Prochaska\altaffilmark{1}, 
Joshua S. Bloom\altaffilmark{2,3},
Hsiao-Wen Chen\altaffilmark{4,5},
Kevin C. Hurley\altaffilmark{6},
Jason Melbourne\altaffilmark{1},
Alan Dressler\altaffilmark{7},
James R. Graham\altaffilmark{8,9},
David J. Osip\altaffilmark{10},
W.D. Vacca\altaffilmark{11,9}
}

\begin{abstract} 

We present Keck/NIRSPEC near-IR images and Magellan/IMACS optical
spectroscopy of the host galaxy of GRB~031203.  The host is an
actively star-forming galaxy at $z=0.1055 \pm 0.0001$. 
This is the lowest redshift GRB to-date, aside from GRB~980425. 
From the hydrogen Balmer lines, we infer an extinction of $A_V
= 3.62 \pm 0.25$ or a total reddening $E_T(B-V) = 1.17 \pm 0.1$ 
toward the sightline to the nebular regions. 
After correcting for reddening, we perform an emission-line analysis and
derive an ISM temperature of $T=13400 \pm 2000$\,K and electron
density of $n_e = 300 \cm{-3}$. These imply a metallicity $\lbrack$O/H$\rbrack$~$=
-0.72 \pm 0.15$\,dex and a roughly solar abundance pattern for N, Ne,
S, and Ar. Integrating \halph, we infer a dust-corrected star
formation rate (SFR) of $> 11 {\rm M_\odot \, yr^{-1}}$. These
observations have the following implications: (1) the galaxy has a
low $K'$-band luminosity $L \approx L_K^*/5$, typical of GRB host
galaxies; (2) the low redshift indicates GRB~031203 had an
isotropic-equivalent $\gamma$-ray energy release smaller than all
previous confirmed GRB events.  The burst discovery, near the
detection limit of INTEGRAL, raises the likelihood of
identifying many additional low $z$, low flux events with {\it Swift};
(3) the large SFR, low metallicity, and the inferred hard radiation
field is suggestive of massive star formation, supporting the
collapsar model; 
(4) several lines of evidence argue against the
identification of GRB~031203 as an X-ray flash event.


\end{abstract}
]

\altaffiltext{1}{UCO/Lick Observatory, University of California, Santa Cruz,
Santa Cruz, CA 95064; xavier@ucolick.org}
\altaffiltext{2}{Harvard-Smithsonian Center for Astrophysics, MC 20, 
60 Garden Street, Cambridge, MA 02138}
\altaffiltext{3}{Harvard Society of Fellows, 78 Mount Auburn Street, 
Cambridge, MA 02138}
\altaffiltext{4}{Center for Space Research, Massachusetts Institute of Technology, 
70 Vassar Street, Building 37, Cambridge, MA 02139-4307}
\altaffiltext{5}{Hubble Fellow}
\altaffiltext{6}{University of California at Berkeley, Space Sciences Laboratory, 
Berkeley, CA 94720-7450}
\altaffiltext{7}{Carnegie Observatories, Carnegie Institution of Washington, 
813 Santa Barbara Street, Pasadena, CA 91101}
\altaffiltext{8}{Department of Astronomy, University of California, 
601 Campbell Hall, Berkeley, CA 94720-3411}
\altaffiltext{9}{Visiting Astronomer, W.M. Keck Telescope.
The Keck Observatory is a joint facility of the University
of California and the California Institute of Technology.}
\altaffiltext{10}{Las Campanas Observatory, Carnegie Observatories, 
Casilla 601, La Serena, Chile}
\altaffiltext{11}{NASA/AMES Research Center, Moffett Field, CA 94035}

\pagestyle{fancyplain}
\lhead[\fancyplain{}{\thepage}]{\fancyplain{}{PROCHASKA ET AL.}}
\rhead[\fancyplain{}{The Host Galaxy of GRB~031203: }]{\fancyplain{}{\thepage}}
\setlength{\headrulewidth=0pt}
\cfoot{}

\section{INTRODUCTION}
\label{sec-intro}

    The study of the host galaxies of gamma-ray bursts (GRBs) 
plays a central role for studies of
progenitor theory and afterglow observations.  First, host 
redshifts are required to derive the burst energy and afterglow timescale
and to determine the 
GRB rate density evolution (e.g. \citealt{djk03}).  Second, the host 
photometric properties \citep{sokolov01,cba02} together with burst locations 
\citep{bloom02} within hosts support the notion of a progenitor population 
intimately connected with star formation (e.g. \citealt{pac98}).  
It is now widely accepted that long-duration
GRBs originate from the deaths of massive stars, as
as suggested by \cite{woosley93} and confirmed by recent observations
\citep[e.g.][]{bloom99,stanek03,hjorth03}.  Third, limits
on time-variable hydrogen- and metal-absorption in the hosts provide 
constraints on the physical state and chemical composition of the interstellar
medium (ISM) in the vicinity of the GRBs \citep{perna98,draine02,mirabal03}.

    At the same time, long-duration GRBs present an alternative
means to address a number of open questions in different sub-fields of 
extragalactic research related to the formation of massive stars.  For example,
early-time spectroscopy of GRB optical afterglows allows us to measure the 
metallicity and dust content of the progenitor environment through absorption
line studies \citep[e.g.][]{vreeswijk01}.  
Late-time galaxy spectroscopy of the GRB hosts allows us to 
study the ionization state and metallicity of the general ISM properties of 
host galaxies using various spectral diagnostics such as the [Ne\,III]/[O\,II] 
ratio \citep{bloom01}.  
Together these results could impose strong constraints on the stellar initial
mass function (IMF) and chemical enrichment history of the galaxy population 
that hosts GRBs.  Previous work presents some evidence that 
GRBs arise preferentially in 
low-metallicity environments with a top-heavy IMF \citep{savaglio03}, which is
supported by the agreement between the luminosity distribution of GRB hosts
and faint blue galaxies found in the field.  In addition, galaxies selected 
by association with GRBs are less affected by dust than those found in optical 
or sub-millimeter surveys \citep{berger03} and therefore may be adopted to
constrain the amount of obscured star-formation in the universe \citep{blain00,
djk03}.  Finally, afterglow studies of GRBs discovered in the early universe
($z \gtrsim 8$) will undoubtedly advance our understanding of the first 
generations of stars as well as chemical evolution in the early universe.

    In this paper we report the spectroscopic identification of the host of 
GRB~031203 at $z=0.1055$ and present a case study of imaging and spectral 
properties of the host population.  The long-duration (20\,sec) GRB~031203 
triggered the IBIS instrument of INTEGRAL on 3 December 2003 at 22:01:28 
UT with an initial localization of 2.5\arcmin\  \citep[radius;][]{Gotz03}. A 
56k sec observation with XMM-Newton beginning at 03:52 UT on December 4, 2003
led to the discovery of an X-ray source (hereafter S1), not present in the 
ROSAT point source catalog \citep{Campana03}, near the center of the 
INTEGRAL error circle \citep[$\alpha=08^h 02^m 30^s.19$, $\delta=-39^\circ 
51' 04''.0$, J2000;][]{Santos-Lleo03}.  Source S1 was later reported to have 
faded over the first XMM pointing \citep{Rodriguez-Pascual03} and the refined 
position of S1 \citep{vaughan04} is consistent with a fading radio source
\citep{Frail03,Soderberg03}. 
Moreover, the expanding X-ray dust echo around the afterglow is
consistent with a bright explosive event (in X-rays) at the
position of the galaxy and coincident in time with GRB~031203
\citep{vaughan04}.
The simplest assumption is that both the X-ray and
radio transients are afterglow emission from GRB~031203.
 
    Before the radio transient was found, several attempts were made to
identify the optical afterglow \citep[see][]{cobb04}.  
The identification of the optical
transient was first reported by \citet{Hsia03} to be within the error circle
of S1, but this identification was invalidated later based on the detection
of sources in the J- and F-plates of the DSS \citep{Bloom03}.  Since the radio 
transient was later found to be coincident with this source and because the
source appeared extended in an $I$-band image, \citet{Bloom03a} suggested that 
the source was a galaxy---either in the foreground or the host of GRB~031203.
Subsequent radio astrometry improved the coincidence of the radio and optical 
source \citep{Soderberg03}, confirming that this galaxy was indeed the host of 
the GRB \citep{pro03a}.  A detailed overview of the optical astrometry of the
host is presented in \cite{cobb04}.  We acquired optical spectroscopy of the 
galaxy on the Magellan telescopes and reported a preliminary redshift of $z=
0.105$ \citep{pro03b}.  Although the host galaxy lies at low Galactic latitude 
and is subject to significant extinction, its proximity affords a careful 
examination of its properties.

    We present detailed IR imaging and optical spectroscopy of the galaxy and
analyze these observations to determine its luminosity, metallicity, and star 
formation rate.  This paper is organized as follows: $\S$~2 reviews the 
observations and data analysis; we perform an emission line analysis in $\S$~3 
to determine the reddening, metallicity, and relative abundances of the
nebular region; and $\S$~4 discusses the star formation rate and implications 
of the galaxy properties for the GRB phenomenon.  Throughout this paper we will
adopt a standard $\Lambda$~cosmology ($H_0 = 70 \mkms$ Mpc$^{-1}$, 
$\Omega_\Lambda = 0.7$ and $\Omega_m = 0.3$).

\begin{figure}[ht]
\begin{center}
\includegraphics[width=3.6in]{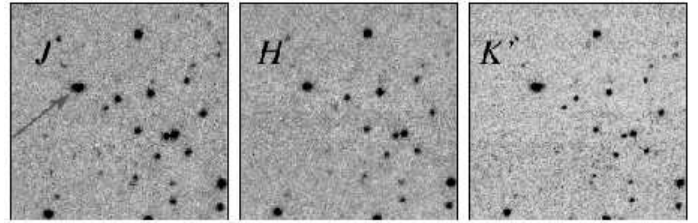}
\figcaption{SCAM images obtained with the NIRSPEC instrument on the
Keck~II telescope on 2003 December 05.6.  
The arrow designates the host galaxy studied in this paper.  
The field of view is $\approx 46''\times 46''$
with orientation N up and E left and the pixel size is 0.183$''$ on a side.
}
\label{fig:img}
\end{center}
\end{figure}

\begin{figure*}
\begin{center}
\includegraphics[width=6.8in]{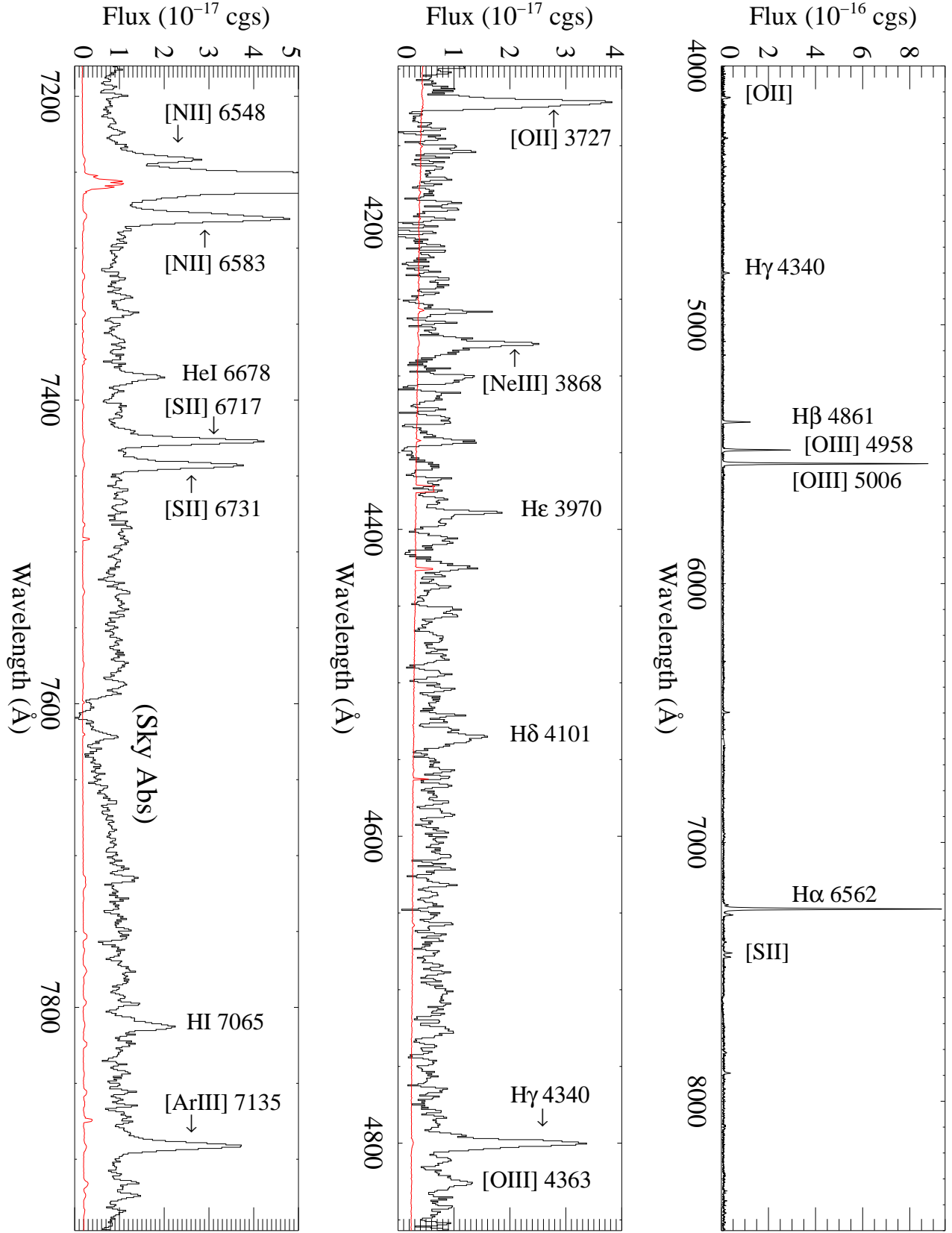}
\figcaption{Magellan/IMACS spectrum of the host galaxy of GRB~031203.
The spectrum has a FWHM~$\approx 5$\AA\ resolution.  The lower
two panels have been smoothed by 3~pix ($\approx$ 1/2 resolution element)
for presentation purposes only.
The strong emission line features are centered at $z=0.1055 \pm 0.0001$
and provide measurements of the extinction and the nebular temperature,
density, metallicity and SFR.  Note the detections of [NeIII] $\lambda 3868$
and [OIII] $\lambda 4363$ as well as the low [OIII]/\hbeta\ ratio.
Together these observations imply a relatively hard radiation field
dominated by massive stars and a metal-poor gas.  The strong \halph\
emission corresponds to a SFR of $> 10 \msol {\rm yr^{-1}}$.
}
\label{fig:spec}
\end{center}
\end{figure*}

\section{OBSERVATIONS AND DATA ANALYSIS}
\label{sec-obs}

  We first observed the galaxy coincident with GRB~031203, host galaxy \grbnm,
on UT 2003 December 05.6 as part of on ongoing 
target-of-opportunity program with the Keck telescope to study GRB afterglows 
and GRB host galaxies.  We imaged the field using the slit viewing camera 
(SCAM) of NIRSPEC \citep{mclean00} in the N3 ($\sim J$-band), N5 ($\sim 
H$-band) and $K'$ filters for total exposure times of 540s each.  The 
observations in each bandpass were composed of nine sets of two to six 
exposures, 10-30 seconds in duration with dither offsets of between 6$''$ and 
9$''$ in space to remove hot pixels.  The coadded images are displayed in 
Figure~\ref{fig:img}.  Although the images were acquired at large airmass 
owing to the low declination of the field, the seeing ranges from FWHM$~\approx 
0.65''$ to $0.5''$ from $J$ to $K'$.  The images clearly identify a galaxy whose 
center lies within $1''$ of the nominal centroid of the radio position for 
GRB~031203 \citep{Soderberg03}.  An elliptical fit to the isophotes yields a PA
angle of $-8$~degrees and an ellipticity $\epsilon = 0.23$.

  The images were calibrated with observations of the Persson standard star 
SJ~9132 \citep{persson98} and we believe the conditions were photometric.  The 
detection limit of the $K'$ image is 20.2 over a $1''$-diameter aperture at the
$5\sigma$ significance level.  The galaxy has $J=18.28$, $H=17.79$, and $K'=
16.54$.  Errors in the flux measurements are $\approx 0.02$\,mag and are 
dominated by systematics such as bias subtraction and aperture correction.

  The following night (2003 December 06.3) we obtained a spectrum of \grbnm\ 
using the recently commissioned IMACS spectrometer \citep{bigelow02} 
on the Magellan/Baade
6.5\,m telescope.  We acquired three exposures totaling 2400s during twilight 
using the 300~line grating centered at 6600 \AA\ with a 0.75$''$ slit.  In this
mode, the instrument delivers a 7~pixel resolution element of ${\rm FWHM}
\approx 5$ \AA.  The data were acquired near transit at an airmass $\approx 
1.02$ and we expect slit losses due to atmospheric dispersion are minimal.
The data are wavelength calibrated and the final wavelength array is converted
to a vacuum scale and corrected to the helio-centric rest-frame.  Finally, we 
calibrate the observed fluxes using a spectrophotometric standard (LTT~9491) 
taken during the start of this photometric night.  We expect a relative flux 
error of $< 10\%$ from the calibration of the spectrophotometric standard, but 
the absolute flux of the GRB host is presumably underestimated owing to slit 
losses (i.e.\ the galaxy overfilled the $0.75''$ slit).

\begin{table*}\footnotesize
\begin{center}
\caption{{\sc EMISSION LINE SUMMARY \label{tab:emlin}}}
\begin{tabular}{lccccccc}
\tableline
\tableline
Ion &$\lambda_{\rm rest}$ & $\lambda_{\rm obs}$ & F$_\lambda$& 
EW$_{\rm rest}$ & F$_\lambda$/F$_{\rm H\beta}$ & L$_{\rm ion}$ \\ 
&(\AA) & (\AA) & ($10^{-16}$ cgs) & (\AA) & & (10$^{41}$ cgs) \\
\tableline\tskip
$[$OII]&3727.26& 4121.96&$   1.57 \pm   0.13 $&$   38.14 \pm   2.02 $&$  1.0607 $&$    5.37$\\
$[$NeIII]&3869.84& 4278.95&$   1.01 \pm   0.09 $&$   28.03 \pm   1.68 $&$  0.6043 $&$    3.06$\\
$[$NeIII]&3970.00& 4388.82&$   0.43 \pm   0.08 $&$   16.33 \pm   1.31 $&$  0.2326 $&$    1.18$\\
H$\delta$&4102.90& 4535.43&$   0.53 \pm   0.08 $&$   21.05 \pm   1.22 $&$  0.2493 $&$    1.26$\\
H$\gamma$&4341.69& 4799.94&$   1.39 \pm   0.07 $&$   32.22 \pm   1.03 $&$  0.4946 $&$    2.50$\\
$[$OIII]&4364.44& 4825.86&$   0.32 \pm   0.06 $&$   12.96 \pm   0.87 $&$  0.1107 $&$    0.56$\\
H$\beta$&4862.70& 5375.60&$   5.18 \pm   0.08 $&$   90.72 \pm   1.10 $&$  1.0000 $&$    5.06$\\
$[$OIII]&4960.29& 5483.52&$  12.07 \pm   0.11 $&$  195.15 \pm   1.56 $&$  2.1108 $&$   10.68$\\
$[$OIII]&5008.24& 5536.50&$  38.09 \pm   0.19 $&$  585.04 \pm   2.76 $&$  6.3565 $&$   32.15$\\
HeI&5877.28& 6497.42&$   1.37 \pm   0.06 $&$   37.24 \pm   0.79 $&$  0.1199 $&$    0.61$\\
$[$OI]&6302.04& 6966.53&$   0.35 \pm   0.05 $&$   15.55 \pm   0.56 $&$  0.0244 $&$    0.12$\\
$[$SIII]&6312.75& 6979.26&$   0.24 \pm   0.05 $&$   13.61 \pm   0.55 $&$  0.0165 $&$    0.08$\\
$[$NII]&6549.91& 7241.21&$   1.03 \pm   0.05 $&$   22.03 \pm   0.61 $&$  0.0630 $&$    0.32$\\
H$\alpha$&6564.63& 7256.84&$  46.32 \pm   0.20 $&$  550.74 \pm   2.31 $&$  2.8167 $&$   14.25$\\
$[$NII]&6585.42& 7279.71&$   2.52 \pm   0.08 $&$   47.78 \pm   0.84 $&$  0.1514 $&$    0.77$\\
$[$SII]&6718.95& 7426.78&$   1.52 \pm   0.06 $&$   27.65 \pm   0.56 $&$  0.0853 $&$    0.43$\\
$[$SII]&6733.16& 7442.63&$   1.28 \pm   0.05 $&$   23.41 \pm   0.53 $&$  0.0714 $&$    0.36$\\
$[$ArIII]&7136.97& 7890.79&$   1.50 \pm   0.08 $&$   34.64 \pm   0.80 $&$  0.0681 $&$    0.34$\\
\tableline
\end{tabular}
\end{center}
\tablecomments{Observed wavelengths in column 3 have been air-to-vacuum corrected and are heliocentric.  The flux $F_\lambda$ and equivalent width
 values are as observed while the flux ratios $F_\lambda/F_{\rm H\beta}$ and luminosities $L_{ion}$ have been corrected for extinction assuming $E_T(B-V) = 1.17$ (see text).}
\end{table*}
 
  The coadded spectrum is presented in Figure~\ref{fig:spec}.  The galaxy shows
several strong emission lines  -- \halph, \hbeta, [OIII], [SII] -- as well as a
series of positive detections including the [OII] doublet (unresolved), 
\hgamma, \hdelta, [OIII]~4363, and [NeIII]~3970.  For each line, we fit a 
Gaussian to the emission line profile to measure the line centroid.  From the 
five strongest emission features, we derive a redshift, $z=0.1055 \pm 0.0001$,
where the uncertainty is dominated by systematic uncertainty (i.e.\ the line 
profile is not strictly Gaussian).  At this redshift, the luminosity distance 
to the galaxy is $487.4\,h_{70}^{-1}$ Mpc, where $h_{70}=H_0/(70 \mkms
{\rm Mpc}^{-1})$, and 1$''$ represents 1.93 $h_{70}^{-1}$ kpc in 
projection.  This measurement marks GRB~031203 as the lowest redshift event to 
date aside from the anomalous GRB~980425 \citep{pian00}.  We also measure 
the emission line-width to be $\sigma = 95 \pm 15 \mkms$ which is consistent 
with the instrumental resolution.  Finally, we sum the line flux and 
equivalent width off the local continuum; these values and the statistical 
errors are listed in Table~\ref{tab:emlin}.

\section{EMISSION LINE DIAGNOSTICS}
\label{sec-emiss}

  In this section, we examine the observed emission line ratios to derive
integrated physical characteristics of the nebular regions within the host
galaxy HG\,031203.  We first estimate the extinction correction necessary for
deriving total line fluxes of the observed emission line.  \grbnm\ is located 
$\approx 4.7^\circ$ off the Galactic plane and has an inferred reddening value 
$E_{G}(B-V) \approx 1.04$ \citep{schlegel98}.  Furthermore, there may be dust 
along the sightline in the emission line region of the host galaxy.  Therefore,
the observed line ratios must be corrected for reddening.  Following 
standard practice, we assess the extinction through comparisons of the 
Balmer line ratios under the assumption of Case~B recombination 
\citep[e.g.][]{osterbrock89, izotov94} and a Galactic extinction law 
$A_\lambda/A_V$ which is parameterized by $R_V \equiv A_V / E(B-V)$ 
\citep{cardelli89}.  Assuming $R_V = 3.1$ for both our Galaxy and the sightline
through \grbnm\ and also adopting a single extinction law\footnote{The 
redshift of \grbnm\, which differs from $z=0$, implies that the extinctions are
not strictly additive; however, this has a minor effect on the total extinction
analysis.}, we derive $A_V = 3.62 \pm 0.25$ which gives a self-consistent 
solution to the relative line strengths of \halph, \hbeta, \hgamma, and 
\hdelta\ and implies a total $E_T(B-V) = 1.17 \pm 0.1$. 
Although this value is consistent with the 
far-IR measurement reported by \cite{schlegel98}, several authors
have argued that the 
far-IR analysis overestimates the $E_G(B-V)$ value
by a factor of $\approx 30\%$ \citep[e.g.][]{dutra03}.  
The Dutra et al.\ analysis recommends scaling the \cite{schlegel98}
value by 0.75 
and therefore one calculates a value of $E_G(B-V) = 0.78$
for the sightline to \grbnm.  In this case, it is likely $(>3\sigma$)
that there is modest extinction $(A_V \approx 1$) associated with the host galaxy.
On the other hand, it is possible that systematic errors in the
Schlegel et al.\ analysis could lead to an underestimate of the reddening
(Finkbeiner 2004, priv.\ comm.).
In the following analysis, we adopt a Galactic reddening $E_G(B-V) = 0.78$ 
and note this may be considered a lower limit.
Of course, the emission line analysis is only sensitive to the total
reddening $E_T(B-V)$ 
and we consider this well constrained for our adopted $R_V$ value
and extinction law.
The extinction corrected flux ratios and luminosities are presented 
in columns 5 and 6 of Table~\ref{tab:emlin}.
We note that a more comprehensive analysis of the differential
extinction (e.g.\ the introduction of multiple extinction laws, variations
in $R_V$) lies both beyond the scope of this paper and dataset, and such
an analysis would
be sensitive to additional uncertainties including absorption from stars
within \grbnm\ and the exact temperature of the nebular regions.

  Next, we assess the physical origin of the emission lines, whether they arise
in an H\,II region, as a result of an active galactic nucleus (AGN), or both.  
A comparison of the measured [OIII]/\hbeta, [SII]/\halph, [OII]/\hbeta, and 
[OI]/\halph\ ratios to those from samples of HII galaxies and AGN presented by 
\cite{rola97} and \cite{kennicutt92} indicate the spectrum of \grbnm\ is
dominated by nebular emission.  
Specifically, AGN show high [S\,II]~$\lambda\lambda 6718,6733$/\halph\ 
and [N\,II]~$\lambda 6583$/\halph\ flux
ratios relative to [O\,III]~$\lambda 5008$/\hbeta\ whereas \grbnm\
exhibits log([S\,II]~$\lambda\lambda 6718,6733$/\halph)$~ = -1.25$ and
log([N\,II]~$\lambda 6583$/\halph)$~ = -1.27$.
These ratios are similar to those 
generally associated with starbursting galaxies, indicating the spectral
features are indeed diagnostics of the star-forming regions that are intimately
connected to the GRB progenitor environment.  Finally, as noted above,
the lines are not significantly broadened but instead show 
a line-width consistent with the instrumental resolution.

  Next, we determine various physical parameters of the star-forming region in 
\grbnm.  To gauge the thermal conditions of the nebular gas, we implement the
IRAF package NEBULAR \citep{shaw95} assuming a two-zone model (low and moderate
temperature).  The [SII] $\lambda\lambda 6717, 6731$ doublet provides an 
assessment of the electron density in the low temperature zone and the relative
emission line strengths of the [OIII] lines yields a measurement of the 
temperature in the moderate temperature zone.  We find $n_e \approx 300 
\cm{-3}$ and adopt this value for both zones.  We then derive a temperature for
the moderate zone $T_{med} = 13400 \pm 2000$K and we assume $T_{low} = 12900$K
based on the prescription suggested by \cite{skillman94}.  The electron
density is well 
constrained by our observations, while the temperatures are less certain owing 
to uncertainties in extinction and the lower signal-to-noise ration of the 
[O\,III] $\lambda 4363$ line.  Allowing for these uncertainties, we derive an 
oxygen abundance [O/H]~$= -0.72 \pm 0.15$~dex (90$\%$ c.l.) assuming $\log 
({\rm O/H})_\odot +12 = 8.74$ \citep{holweger01}.  

\begin{figure}[ht]
\begin{center}
\includegraphics[height=3.6in,angle=90]{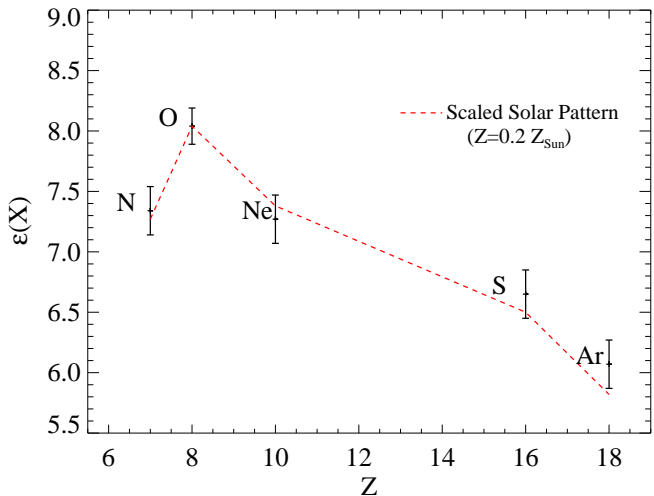}
\figcaption{Abundance pattern of the emission-line regions of 
\grbnm.  The dashed line traces the Solar abundance pattern
which has been scaled to the galaxy's oxygen 
abundance (log(O/H) + 12 = 8.1).  We find no significant deviations
from the solar pattern for \grbnm.
}
\label{fig:solabd}
\end{center}
\end{figure}

  Finally, we measure relative elemental abundances for N, Ar, Ne, and S after 
adopting the ionization corrections advocated by \cite{izotov94} and solar 
relative abundances \citep{grvss96,holweger01}.  We find: [N/O]~=$+0.07$, 
[Ne/O]~$=-0.11$, [S/O]~$=+0.15$, and [Ar/O]~$=+0.25$.  The uncertainty in these
values is $\approx 0.2$~dex and all of the values are consistent with a solar 
pattern (Figure~\ref{fig:solabd}).  This matches our expectation for the various 
$\alpha$-elements and the measurements further support a low metallicity for 
\grbnm.  We note, however, that the [N/O] value is $\approx 2\times$ larger 
than expected given the sub-solar metallicity \citep[e.g.][]{henry00}.  It is 
possible that we are underestimating the continuum level near the [N\,II] 
$\lambda\lambda 6544, 6585$ lines owing to the very strong \halph\ line and 
therefore are overestimating the N abundance.

\section{DISCUSSION}

\subsection{Luminosity, Metallicity, and Star Formation Rate}

Even with nearly one magnitude of foreground extinction at 1$\mu$m, 
\grbnm\ offers a rare opportunity to study the large-scale environment
of a nearby GRB event. 
Our analysis of the nebular region
of \grbnm\ reveals a metal-poor, star-forming galaxy.  
We can combine our extinction analysis with the apparent near-IR magnitudes
and the redshift of \grbnm\ to assess the luminosity of the
old stellar population within the galaxy.
Adopting a Galactic reddening of
$E_G(B-V) =  0.78$ (see above), 
we derive extinction corrections of 0.67, 0.45, and 0.28 magnitudes for
the $J, H$ and $K'$ bands respectively.  
Adopting the standard $\Lambda$~cosmology, $z_{gal} = 0.1055$ 
and a k-correction of $-0.2$~mag,
we calculate absolute magnitudes of $-21.20$, $-21.47$, and
$-22.35$~mag in $JHK'$ respectively.
The resulting $K'$ luminosity is +1.9~mag fainter than $M(K)_*$ at
$z=0$ \citep{cole01}, i.e.,  
the galaxy has a low luminosity $(L \approx L_K^*/5)$ 
and presumably a low total mass.
In these respects, \grbnm\ is a relatively ordinary, faint galaxy.
The $K'$-band luminosity is consistent with the median absolute K magnitude
compiled by \cite{lefoch03} for GRB host galaxies.  Apparently, these
galaxies are, as a class, underluminous in terms of their old stellar
population.

The galaxy is notable, however, for a large star formation rate (SFR).
With our spectroscopic observations, we infer the SFR from the integrated,
extinction-corrected \halph\ luminosity (Table~\ref{tab:emlin}).   
Adopting the \halph\ relation given by 
\cite{kennicutt98},we find SFR(\halph)~$= 11 \, {\rm M_\odot yr^{-1}}$. 
We believe the uncertainty in SFR(\halph) is dominated by uncertainties in 
the SFR calibration which \cite{kennicutt98} estimates to be $\approx 30\%$.
We also attempt to infer the SFR from the dust-corrected
[OII] luminosity using the comprehensive prescription 
described by \cite{kewley04}.  
Unfortunately, their analysis is only applicable to emission-line regions
with log(O/H)~+12~$> 8.2$ (Kewley, priv.\ comm.);  the SFR corrections
diverge at the nominal metallicity of \grbnm\ 
(see their Figure~9)\footnote{These points not withstanding,
we caution the reader that SFR values, irrespective of reddening corrections,
derived for other GRB hosts based 
on the [OII] lines alone may have uncertainties of $>100\%$, especially if 
the metallicity is poorly constrained.}.
In addition, the [OII] lines
have greater sensitivity to the dust corrections and much poorer SNR than
the \halph\ emission.
Therefore, we consider only the SFR value implied by \halph.
We stress that this SFR value should be considered a lower limit to the
total SFR because (i) the galaxy is larger than the slit used and
(ii) the galaxy may contain regions of star formation which are
enshrouded in dust.
The latter effect is presumably small for this relatively metal-poor
galaxy while the former effect may increase the SFR by up to 50$\%$.
Indeed it will be of interest to measure the late-time radio and
sub-mm flux from this galaxy to assess the level of obscured star
formation and compare long-wavelength emission of this host with GRB
hosts at higher redshifts \citep[following][]{berger03}.


To qualitatively assess the nature of star formation
for \grbnm, we would like to 
contrast its characteristics with local samples.
One simple comparison is the rest-frame equivalent width of the
\halph\ line.  For \grbnm, we measure 
EW$_{\rm rest}$(\halph)~$= 550.7  \pm 2.3$\AA\ which is significantly
larger than normal, star-forming galaxies \citep{jansen00,nakamura04}.
This large value suggests a system undergoing a short, very intense burst
of star formation.

Another valuable diagnostic is the ratio of SFR to total luminosity.
Unfortunately, there is no large, single survey to date which has 
compared SFR and near-IR luminosity.  We therefore estimate the
B-band luminosity from the measured continuum flux 
at $\lambda_{obs} \approx 4500$\AA.  We observe 
$F_{4500} = 8.0 \sci{-18}$~ergs/s/cm$^{-2}$/\AA\ which translates to a Vega
magnitude $B=18.1$\,mag, consistent with the detection of the
galaxy in the DSS-J plate \citep{Bloom03}.  
Adopting $E_G(B-V)=0.78$, we impose a dust correction
of 3.1~mag and derive an absolute magnitude $M_B = -19.3$.  
We stress that this luminosity is likely an underestimate of the total
B-band light because the galaxy overfilled the 0.75$''$ slit.
Furthermore, this $E_G(B-V)$ value may be an underestimate
(see the discussion in $\S$ref{sec-emiss}).
We compare the measured \halph/B-band luminosity of \grbnm\ against 
the KPNO International Spectroscopic Survey, an emission-line survey
of galaxies with $z<0.095$ selected in low-dispersion objective-prism
spectra \citep{salzer00}.
Restricting our comparison to those 
galaxies with accurate measurements of $L({\rm H\alpha})$ and 
$M_B$ \citep{gronwall04}, we
note \grbnm\ falls at the upper end of the distribution, i.e. its SFR is
$\approx 5 \times$ higher than galaxies with similar B-band luminosity.
A portion of this offset could be explained by correspondingly
higher slit losses for the B-band light than \halph\ and/or a higher
$E_G(B-V)$ value.  
These corrections not-withstanding, 
we suspect the galaxy has a higher than average
SFR per unit B-band luminosity than other local galaxies.

We can also place \grbnm\ on the metallicity/luminosity
locus of KISS galaxies \citep{melbourne02}.  \grbnm\ lies below the
entire distribution.  Furthermore, it falls
$\Delta{\rm (O/H)} \approx -0.9$~dex or $\Delta M_B \approx -2$~mag
off their fit to the KISS sample even if we adopt $M_B = -19.3$\,mag.
We speculate this offset is characteristic
of a very young star forming region.  Perhaps
we have observed the galaxy prior
to the production and/or distribution of significant metals into the
nebular regions.  Indeed, a similar effect is observed for the star-bursting
Lyman break galaxies at high redshift \citep{pettini01,shapley04}. 

\subsection{The case for identifying \grbnm\ as the host galaxy of
GRB~031203}

The conclusions drawn in the following subsections hinge on the 
identification of \grbnm\ as the host galaxy of GRB~031203. 
In particular, this sets the redshift of GRB~031203 based on the
observed emission lines of \grbnm.
Before proceeding, therefore, we review the evidence for this
allegation.  First, \cite{soderberg04} and \cite{Rodriguez-Pascual03} 
respectively have localized fading radio and x-ray sources 
within the half-light radius of \grbnm\ \citep[see][]{cobb04}.  
Integrating the K-band number density function
of \cite{chen04} to $K'=16.5$, the number density of galaxies 
is $7.7\sci{-5} / \square''$.  Therefore, the likelihood of laying within $2''$
of a galaxy at least as bright as \grbnm\ is $<0.1\%$.  
Second, \grbnm\ has a $K'$-band luminosity typical
of the GRB host galaxy distribution \citep{lefoch03}.

Third, this galaxy exhibits a SFR $(>10 \msol {\rm yr^{-1}})$
which is $> 98\%$ of all galaxies at $z \lesssim 0.1$ \citep{nakamura04}.
This argument is independent of the apparent magnitude and, furthermore,
we note that the SFR is comparable to that of other GRB hosts 
\citep[e.g.][]{berger03}.  Fourth, the galaxy is metal-poor, a trait
frequenly attributed to GRB host galaxies.
Fourth, \grbnm\ exhibits a lower metallicity than $>99\%$ of all galaxies
with $M_B < -19.3$\,mag \citep{melbourne02,lamareille04}.
Again, GRB hosts appear to be metal-poor in general.  
Finally, there is recent evidence that a supernova was associated with
GRB~031203 with a location coincident with \grbnm\ 
\citep{thomsen04,cobb04,galyam04}.
Although none of these points can be considered a definitive argument
for associating \grbnm\ with GRB~031203, together they present a very
strong case for a physical connection between GRB~031203 and \grbnm.

\subsection{The Energetics of GRB~031203 and Arguments against an X-ray Flash}
\label{sec:ibis}

The fluence in the prompt gamma-ray emission in GRB~031203 has not yet been
reported. To estimate this quantity, we fit the INTEGRAL SPI-ACS
light curve\footnotemark\footnotetext{{\tt
http://isdc.unige.ch/cgi-bin/cgiwrap/$\sim$beck/ibas/spiacs/ibas\_acs\_web.cgi}}
with a double exponential (single pulse).
Assuming Poisson weighting, scaling the reported peak flux to
1.3 $\times 10^{-7}$ erg cm$^{-2}$ s$^{-1}$ (20--200 keV), and
assuming that the IBIS {\it count} rate is dominated by the photons in
this low energy range, we estimate the fluence to be
$(4 \pm 1) \times 10^{-7}$ erg cm$^{-1}$ (20--200 keV). 
This assumes a combined uncertainty of 25\% in the peak flux and
integrated (model) light curve.
We can estimate the prompt isotropic-equivalent energy release
$E_{iso}$(20--2000 keV) by assuming a $k$-correction to rest-frame (20--2000 keV)
from an ensemble of bright bursts: $k = 2.6 \pm 0.9$.
We find $E_{\rm iso}$(20--2000 keV) $= (2.6 \pm 1.1) \times 10^{49}
h_{70}^{-2}$ erg with a plausible maximum of $\approx 5
\times 10^{49}$ erg if the burst was spectrally hard ($E_0 =
1000$ keV). This is nearly identical to an independent estimation by
Watson et al., who found (under a differing set of assumptions) $E_{\rm
iso}$(20--2000 keV) $= 2.6 \times 10^{49} h_{70}^{-2}$ erg
(scaling to our chosen value of Hubble's constant).

As noted in Watson et al. (2004), this value of $E_{\rm iso}$(20 --
2000 keV) is about 30 times fainter than that inferred in the
(geometry corrected) prompt emission of other cosmological GRBs
\citep{bfk03}. If this burst was collimated, then the true energy
release in the $\gamma$-ray bandpass was even lower. Soderberg et
al. (2003) also pointed out that the kinetic energy in the blastwave,
as proxied by the X-ray afterglow emission, was $10^{3}$ times lower
than other GRBs at comparable times (Berger al. 2003). These two
results seem to suggest that the total energy in the relativistic
component was substantially lower than the other cosmological GRBs.

Watson et al.\ (2004) have suggested that GRB~031203 
may be an X-ray flash based upon analysis of the dust echo. However,
since the occurrence of a bright soft X-ray component to the prompt
burst has not been firmly established, we consider the XRF
possibility less likely than Watson et al.\ (2004). 
We note that the Watson et al.\ argument rests on two points.  First, that
the ratio of the 0.2-10 keV {\it fluence} (estimated from the dust
echo) to the 20-200 keV fluence (estimated from the {\it peak flux} 
of the GRB) yields a reliable power law index.  And second, that the
detection of this burst at energies $\gtrsim 100$ keV by the INTEGRAL SPI-ACS
is consistent with the spectral
slope estimate.  We caution however that the first point mixes fluences
and peak fluxes and cannot take any possible spectral evolution into
account; until a reliable time-integrated spectrum above 20 keV
is given, this procedure is subject to considerable uncertainty.  And
contrary to claim of Watson et al.\ we believe that the spectrum derived
from this procedure cannot be said to be consistent with the SPI-ACS
response.  Indeed, the ACS threshold is not well-defined, as it varies
along the collimator, and blocking by INTEGRAL instruments for various
angles complicates the response considerably.  There is at present no
accurate calibration of the ACS.  Thus we believe that it is impossible
to confirm or disprove the X-ray rich burst hypothesis based on either argument.

\cite{vaughan04} have reported a dust scattered X-Ray halo
from {\it XMM-Newton} observations taken six hours after the burst.
Comparing the spectral shape of the halo with the afterglow, they
inferred a hydrogen column density $N_H = 8.8 \pm 0.5 \sci{21} \cm{-2}$
consistent with the Parkes 21cm observations along this sightline
\citep{mcclure01}.
\cite{vaughan04} then estimated a time-integrated X-ray flux by assuming
$A_V = 2$\,mag, a value 4.4$\times$ smaller than out total derived extinction.
We caution, therefore, that they may have overestimated the X-ray flux in
their analysis. 
Thus, we suggest GRB~031203 is most like
GRB~980425, i.e.\ a low-redshift underluminous burst associated with a 
supernova.  In this respect, it is interesting to note
the single pulse nature of the burst \citep{bloom98}, which
may be related to the emission mechanism.

\subsection{Implications for the GRB event}

Collapsars \citep{woosley93}, the leading scenario for long-duration
GRBs, requires a connection between the instantaneous ($\lesssim 10^7$~yr)
star-formation and the probability of bursting \citep{bloom98,fryer99}.
This model describes the GRB progenitor
as a massive star ($> 30 \msol$ at zero-age main sequence) 
which loses all of its 
hydrogen envelope and most of the helium.
Importantly, \cite{macfadyen99} noted that low metallicity favors a
collapsar event because low metallicity decreases mass loss 
leading to a more massive, more rapidly rotating star, i.e.\ characteristics
necessary to make a disk and black hole.
Examining \grbnm\, in this context,
we note: (i) the presence of [NeIII] emission;
(ii) O$^{++}$ is the dominant oxygen ion for the full range of extinction
and temperatures considered; and
(iii) the emission line regions have a significantly sub-solar metallicity.
Points (i),(ii) indicate the presence of a relatively hard radiation
field.  That is, massive stars $(> 30 \msol$) contribute significantly
to the spectral energy distribution of the host.
Furthermore, point (iii) follows the assertion of 
\cite{macfadyen99} that the collapsar event is more easily achieved in lower
metallicity systems.  This point is accentuated by the fact that
the system has a lower than average metallicity given its B-band luminosity.
Together, these observations of \grbnm\ lend support to the collapsar
model for GRB's.

Aside from the anomalous GRB~980425, the only other long duration GRB
with a known low redshift $(z < 0.2)$ is GRB~030329 
\citep[see][for a review]{lipkin03}.
The discovery of GRB~031203 at such a low redshift has several
important implications.  First, it indicates the likelihood of 
many additional low redshift GRB events at low flux/fluence levels.
Since so few low redshift bursts had been
discovered in the past 6 years (1 out of $\sim$100 well-localized
GRBs) and the peak flux of GRB~030329 was more than two orders of
magnitude over the detection threshold, on probabilistic grounds,
\citet{price03} suggested that bursts with redshifts as low as GRB 030329 
would be rare even in the {\it Swift} era (see also Schmidt 1999). 
The low redshift discovery of GRB\,031203, however, requires a redress
of this conclusion. 
Second, the peak flux of 
1.3 $\times$ 10$^{-7}$ erg cm$^{-2}$ s$^{-1}$ \citep[20--200 keV;]{Mereghetti03} 
implies that GRB~031203 was one of the faintest
rapid and well-localized bursts to date. 
While there is still some uncertainty in the
$k$-corrections, it appears that the {\it prompt}
isotropic-equivalent energy of GRB~031203 
appears to bridge the gap between the anomalous 980425 and
the remainder of the bursts (see also Soderberg et al.\ 2003). Again,
we point out that 031203 is distinguished from other
low-energy GRBs (030329, 980326, etc.; Bloom, Frail, \& Kulkarni
2002\nocite{bfk03}) in that the energy of 031203 in soft $\gamma$-rays
(20--2000 keV) appears low even before (the unknown) geometry
correction. Given the uncertainties related to estimations of the
energy release in prompt X-rays, the total energy budget of the
burst is rather uncertain. Analysis of the radio afterglow
\citep{soderberg04} may eventually yield a tighter
constraint on the total kinetic energy of the ejecta.
Lastly, given the low volume associated with the
universe at $z \lesssim 0.1$, one surmises that the frequency distribution
of GRB energies likely increases to lower energy.  
This may be naturally explained as
the effect of viewing angle to the GRB event \citep[e.g.][]{woosley99}, 
but could also be the result of an intrinsically broad distribution 
of GRB luminosities \citep[e.g.][]{kouveliotou04}.
In either case, the discovery of GRB~031203 significantly 
increases the likelihood
of detecting many additional low $z$, low energy GRB events with the launch
of {\it Swift}.

{\it Note Added After Original Submission:} Newly surfaced $\gamma$-ray
satellite data in Sazonov, Lutovinov, \& Sunyaev (2004) (recently
relayed in Gal-Yam et al.~2004), appears to exclude the XRF hypothesis.
This confirms our prior inferences --- that this event was a
low-luminosity GRB and not an XRF --- presented in section 4.3.

\acknowledgments
We acknowledge the excellent efforts of the staff at Keck and Magellan
Observatories who aided in acquiring these observations.
We thank the members of the SMARTS GRB group at Yale for
collaborative access to their imaging data. 
We acknowledge helpful discussions with J. Salzer, S. Woosley, 
B. Gaensler, M. Haverkorn, R. Smith, D. Finkbeiner, and P. van Dokkum.
JSB is grateful for
support from the Harvard Society of Fellows and the
Harvard-Smithsonian Center for Astrophysics.
H.-W.C. acknowledges support by NASA through a Hubble Fellowship grant
HF-01147.01A from the Space Telescope Science Institute, which is operated by
the Association of Universities for Research in Astronomy, Incorporated, under
NASA contract NAS5-26555.
The authors wish to recognize and acknowledge the very significant cultural
role and reverence that the summit of Mauna Kea has always had within the
indigenous Hawaiian community.  We are most fortunate to have the
opportunity to conduct observations from this mountain.


\end{document}